\documentclass[fleqn,10pt, twocolumn]{wlscirep}
\usepackage[utf8]{inputenc}
\usepackage[T1]{fontenc}

\definecolor{airforceblue}{rgb}{0.36, 0.54, 0.66}
\definecolor{brickred}{rgb}{0.8, 0.25, 0.33}
\definecolor{amber}{rgb}{1.0, 0.75, 0.0}
\definecolor{applegreen}{rgb}{0.55, 0.71, 0.0}
\definecolor{magenta}{rgb}{0.965, 0, 0.859}

\usepackage{nameref}

\title{Language bubbles in online social networks}

\author[1,2,3]{Alessandro Bellina}
\author[3,2,1]{Donald Ruggiero Lo Sardo}
\author[3,2,1]{Emanuele Brugnoli}
\author[2,4,5]{Fabio Saracco}
\author[6,2]{Pietro Gravino}
\author[1,3,2,7,*]{Vittorio Loreto}
\author[8,3,2]{Gabriele Di Bona}
\affil[1]{Sapienza University, Physics Dept., P.le A. Moro, 2, I-00185 Rome, Italy}
\affil[2]{Centro Studi e Ricerche Enrico Fermi (CREF), Via Panisperna 89/A, 00184 Rome, Italy}
\affil[3]{Sony Computer Science Laboratories - Rome (Sony CSL-Rome), Joint Initiative CREF-SONY, Via Panisperna 89/A, 00184, Rome, Italy}
\affil[4]{CNR, Institute for Applied Computing ``Mauro Picone'' (IAC), Via Madonna del Piano 10, 50019 Sesto Fiorentino, Italy}
\affil[5]{IMT, School for Advanced Studies Lucca, P.zza San Francesco 19, 55100 Lucca, Italy}
\affil[6]{Sony Computer Science Laboratories - Paris (Sony CSL-Paris), 6, Rue Amyot, 75005 Paris, France}
\affil[7]{Complexity Science Hub Vienna, Metternichgasse 8, 1030 Wien}
\affil[8]{CNRS, GEMASS, 59 rue Pouchet, F-75017, Paris, France}
\affil[*]{vittorio.loreto@roma1.infn.it }

\begin{abstract}
Social media platforms have become essential spaces for public discourse. While political polarisation and limited communication across different groups are widely acknowledged, the connection between social network fragmentation and the language features and quality used by various communities has received insufficient attention. This study aims to fill this gap by examining the social structure and linguistic richness of the Italian debate on Twitter/X. We analyse tweets and retweets from Italian politicians and news outlets between 2018 and 2022, characterising the retweet network and evaluating the language used within different communities through various lexical metrics. Our analysis uncovers two systematic patterns: communities closer in the network tend to use more similar vocabulary, while isolated communities consistently demonstrate lower lexical diversity and richness. Together, these patterns illustrate what we call ``language bubbles''. These findings indicate that socially isolated communities interact less with others and develop distinct and poorer linguistic profiles, highlighting a structural link between social fragmentation and linguistic divergence.
\end{abstract}
\begin{document}

\flushbottom
\maketitle

\thispagestyle{empty}

\section{Introduction}

Online social networks significantly impact how people consume and interact with content~\cite{bakshy2012role, guille2013information, pariser2011filter}. As central hubs for information exchange, these platforms enable the rapid sharing of information, making a variety of perspectives widely available~\cite{castells2011rise, lerman2010information}. However, information consumption often trends towards narrower and more specific content due to human tendencies for homophily~\cite{mcpherson2001birds, bisgin2012study} and algorithmic recommendations aimed at maximising user engagement~\cite{de2022modelling, bellina2023effect}. These mechanisms reinforce users’ preferences, leading to feedback loops of selective exposure. The broader consequences of this phenomenon are still debated, especially regarding how it shapes attitudes and beliefs~\cite{Boxell2017,Budak2024}.

One significant consequence of this dynamic is the emergence of distinct and separated communities within social networks, a phenomenon commonly known as polarisation~\cite{conover2011political, flamino2023political}. This social fragmentation occurs when users form highly segregated groups that share similar perspectives while contrasting sharply with others~\cite{garimella2018political}. This phenomenon has been widely documented across major online platforms~\cite{barbera2015tweeting, cinelli2021echo}, and it is not limited to online environments, as it can also emerge in offline social settings~\cite{sunstein2009going, baldassarri2007dynamics}.

Another closely related phenomenon observed in social platforms is the presence of the so-called ``echo chambers''~\cite{flaxman2016filter, brugnoli2019recu, cinelli2021echo, Pratelli2024b}. The existence of well-defined communities, each with its own set of ideas and opinions, favours the tendency of users to primarily engage with information that confirms their existing beliefs while avoiding opposing perspectives~\cite{cinelli2020sele}. This behaviour reinforces biases and isolates communities, ultimately limiting the diversity and complexity of ideas being shared~\cite{yasseri2016political, bennett2013logic}. 

A related concept is ``filter bubbles'', which refers to personalised information environments shaped by algorithms, user behaviour, and social structures. These environments may reduce users' exposure to diverse viewpoints~\cite{pariser2011filter, michiels2022filter, bellina2023effect}. While the empirical relevance of filter bubbles remains debated~\cite{bruns2019filter}, the term has influenced public and academic discourse on how algorithmic systems might interact with individual preferences to shape information environments. Both echo chambers and filter bubbles have been associated with reductions in informational diversity~\cite{kitchens2020understanding,piao2023human-ai}.

Despite these insights, previous research has seldom measured how factors like social segregation, fragmentation, and the emergence of distinct social niches impact the quality of social discourse by directly analysing language patterns among users. Interestingly, a recent study examined how news sources in the United States, specifically CNN and FOX, assign different meanings to the same words, resulting in a semantic separation in their language~\cite{ding2023same}. The researchers accomplished this by embedding words from news headlines from both sources and tracking how the embeddings diverged over time. Another related study explored the evolution of users' vocabulary size and richness, but it did not explicitly connect these metrics to particular communities, social niches, or patterns of isolation~\cite{di2024evolution}. A complementary study found that politically aligned communities in the U.S. increasingly diverge in word usage and meaning, suggesting that social segregation reshapes not only interaction patterns but also the semantics of everyday language~\cite{karjus2024evolving}. However, it remains unclear to what extent these forms of social segregation correlate with a reduced ability for groups to exchange ideas and develop complex perspectives that incorporate diverse viewpoints~\cite{Bail2018, sunstein2009going, iyengar2015fear}. A comprehensive analysis of these implications is still lacking.

In this study, we explore the relationship between the quality of social discourse—measured through lexical observables—and the levels of social segregation and community fragmentation within the system. We examine various topics in a large dataset from Twitter (now X), which includes millions of tweets and retweets from thousands of Italian information leaders, including politicians and news providers, from 2018 to 2022. Our analysis focuses on the presence and structure of communities by investigating the topology of the retweet network and quantifying the level of segregation among these communities. This approach relies on interpreting retweets as a form of endorsement~\cite{conover2011political, caldarelli2021flow, brugnoli2024fine}, a widely accepted assumption in the literature that helps reveal underlying community structures. To evaluate the quality of social discourse, we analyse the language used by different communities through a set of lexical metrics. These metrics capture the linguistic distance between communities and their lexical richness and diversity.

Our results reveal a phenomenon we call \textit{language bubbles}: structurally segregated communities that not only interact little with each other but also develop increasingly divergent and impoverished linguistic repertoires. Specifically, this phenomenon is characterized by two distinct patterns. On the one hand, communities that are closer in the social network, i.e., those that share a large portion of their retweeters, tend to exhibit more similar linguistic patterns. In contrast, more distant communities show greater divergence. On the other hand, the diversity and complexity of language, as measured by our lexical metrics, systematically decrease in the most isolated and extreme niches of the network, an effect that becomes more pronounced in globally more fragmented environments. Overall, our findings highlight a strong correlation between the retweet network's topology and communities' linguistic characteristics, offering new perspectives on the implications of social segregation and fragmentation on online platforms.

\section{Results}

\label{sec:results}

    \subsection{Identifying discourse communities in the retweet network}

    \label{sec:results_1}

        For our analysis, we utilised a corpus consisting of approximately 14 million tweets from Italian politicians and news outlet accounts, along with their retweeters. The dataset covers the period from 2018 to 2022, prior to \textit{Twitter}'s rebranding as \textit{X}. 
        According to the rating agency NewsGuard, our list of news outlets covers accounts for approximately $95\%$ of online news engagement. This ensures a broad and representative sample of the global news-related conversation. Although the influence of these accounts may vary across specific topics, they remain central for evaluating engagement and endorsement patterns across the public discourse (for further details on the data, see Section~\ref{sec:methods}). From this large corpus, we extracted six datasets related to popular topics, namely \textit{immigration}, \textit{vaccines}, \textit{climate}, \textit{sport}, \textit{music}, and \textit{cars}. These topics were selected to cover a broad spectrum of public debates, ranging from socially and politically sensitive issues to more neutral ones. This allows us to explore how different kinds of discussions---some more contentious, others more consensual---relate to patterns in network structure and language use, as explored in the following sections. We conduct separate analyses for each topic and then compare the results across them.

        To explore the relationship between the social fragmentation and the linguistic characteristics of different topics and communities in the social network, we first define observable metrics that can be measured on the dataset. 
        In our analysis, we assume that each retweet represents a positive interaction between two users. This allows us to first construct a bipartite network for each dataset, where one set of nodes represents the monitored users active in the related topic (i.e., producing at least one tweet retweeted at least once), which we refer to as \textit{influencers}, and the other set of nodes represents the accounts who have retweeted at least a tweet of the influencers in the dataset.

        \begin{figure*}[htb!]
            \centering
            \includegraphics[width=.9\linewidth]{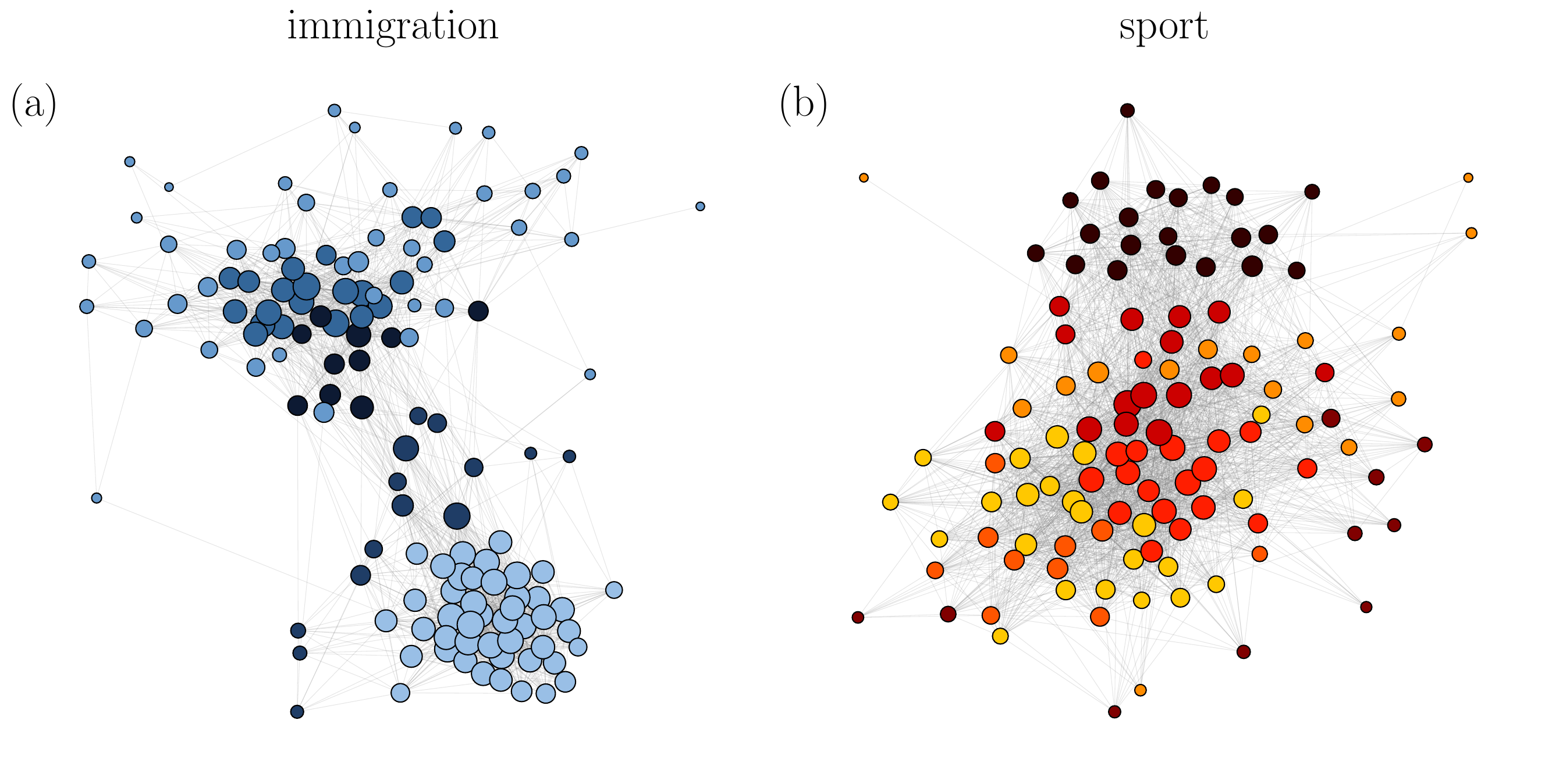}
        
            \caption{\textbf{Retweet network from Twitter data.} 
            Example of a projected monopartite network of influencers for the topics \textit{immigration} \textbf{(a)} and \textit{sport} \textbf{(b)}, based on retweet activity from 2018 to 2022. Nodes represent influencer accounts ($N_{users} = 131$ and $115$, respectively), and their size is proportional to the number of retweets received. Links indicate similarity, measured by the number of accounts that have retweeted both users. Communities, identified using the hSBM algorithm, are shown in different colours ($N_{communities} = 5$ and $7$, respectively). The two networks exhibit markedly different structures: in \textbf{(a)}, communities are more sharply separated, while in \textbf{(b)} they are more interconnected. This difference anticipates the concept of \textit{topic fragmentation}, which is formally defined in Eq.~\ref{eq:D_topic}.}

            \label{fig:fig1}
        \end{figure*}

        By projecting this bipartite network onto the layer of influencers, we obtain a co-occurrence weighted network, where the connection between two influencers represents the number of users who have retweeted both (for details, see Section~\ref{sec:methods}). 
        As an example, Figure \ref{fig:fig1} displays such a co-occurrence network for the topic of \textit{immigration} (panel a) and \textit{sport} (panel b). 
        This projection captures the similarity between accounts in terms of shared audience, enabling us to assess the degree of separation between communities. 
        Using the hierarchical stochastic block model (hSBM)~\cite{peixoto2014hierarchical,peixoto2017nonparametric}, we identify distinct communities of influencers who primarily interact within their own community but rarely engage with members of other groups. 
        
        This community structure allows us to quantify both the pairwise separation between communities and the overall fragmentation of a topic's network. 
        For instance, in the case of \textit{immigration} (a), the community structure appears more sharply defined, with clusters that are clearly separated from each other. Conversely, for \textit{sport} (b), the boundaries between communities are more diffused and the network appears less fragmented. We will later formalise this notion of \textit{topic fragmentation}, showing that the fragmentation level is higher for \textit{immigration} than for \textit{sport}. 

        \subsection{Linguistic alignment across discourse communities}
    \label{sec:results_2}
    
        \begin{figure*}[htb!]
            \centering
            \includegraphics[width=0.9\linewidth]{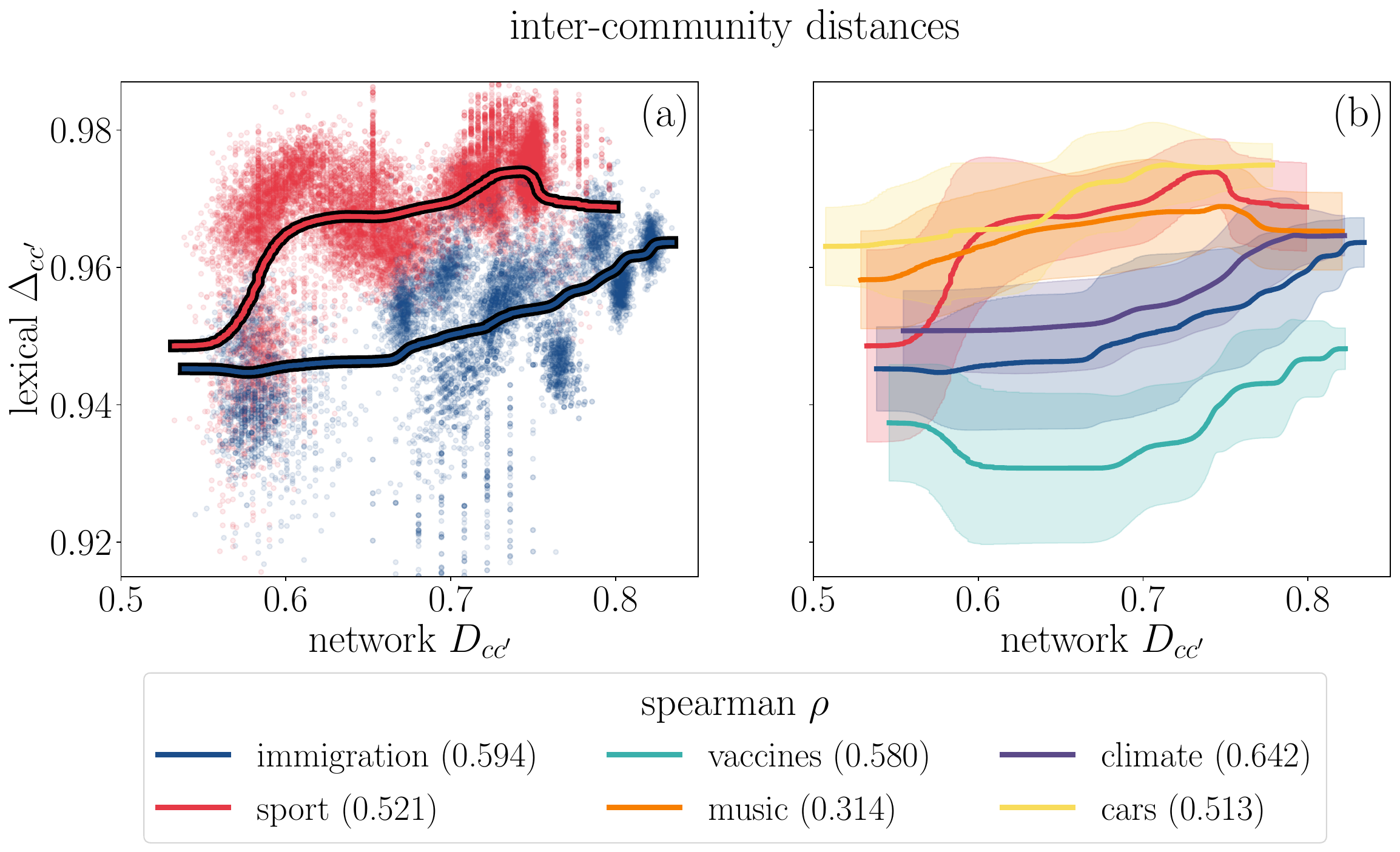}
        
            \caption{\textbf{Lexical divergences across communities correlate with distances on the retweet networks.} \textbf{(a)} Jensen-Shannon divergence of communities as a function of their distance on the retweet network, for the topics \textit{immigration} and \textit{sport}. Each point in the plot represents a pair of communities, for which we compute both the JS divergence between their token distributions and their distance on the retweet network. The plot is generated by subsampling each community 1,000 times, each time randomly selecting 60\% of its members.
            Solid lines indicate the moving average of the median.
            \textbf{(b)} Moving average and confidence interval between the first and third quartile for all six topics. Each colour represents a topic, with warm colours corresponding to neutral topics and cool colours to potentially polarising ones. In all six cases, a significant effect is observed: in general, communities that are farther apart on the retweet network exhibit greater lexical divergence. These observations reveal a first manifestation of the broader phenomenon of language bubbles. The statistical significance of the results is tested against null models (see Table~\ref{tab:null_hsbm}). 
            }
            \label{fig:fig2}
        \end{figure*}

        After defining the network of similarities between influencers and identifying the discourse communities for each topic, we can measure their separation both in terms of network structure and language use. We first define the \textit{inter-community network distance} $D_{cc'}$ between two communities $c$ and $c'$ as the average shortest-path distance between all pairs of nodes belonging to the two communities in the retweet network (see Eq.~\ref{eq:D_cc'} and Section~\ref{sec:methods} for details).
        
        In parallel, we investigate differences in language use by computing the \textit{inter-community lexical distance} $\Delta_{cc'}$, defined as the average lexical distance between all pairs of influencers from the two communities (Eq.~\ref{eq:Delta_cc'}). Given the frequency distribution of words used by each influencer, we quantify their lexical distance using the Jensen-Shannon Distance~\cite{cover1999elements} (JSD), a standard information-theoretic measure that evaluates how much two probability distributions differ. A JSD of 0 indicates identical distributions, while higher values (bounded above by 1) reflect greater dissimilarity in lexical usage (see Section~\ref{sec:methods} for implementation details).
        Notice that in our pipeline, we first remove the top 1000 most frequent tokens across all topics, i.e., we remove the \textit{kernel lexicon}~\cite{cancho2001small, bellina2025cognitive}, which represents the portion of language commonly shared by all speakers and is not indicative of individual differences. For this reason, the frequency distribution of words for each influencer is less dense than normal, thus obtaining generally high values of JSD in our following analysis (see Appendix~\ref{app:kernel} and Appendix~\ref{app:test_metrics} for further details).
                
        Having defined a network and lexical inter-community distance, we first explore how lexical differences manifest within the retweet network. More specifically, we examine whether distances in the retweet network correlate with linguistic divergences among discourse communities in each topic. By analysing these two independently computed quantities---one based on the retweet network structure and the other one on the content of user-generated tweets---we can determine whether a relationship exists between them. 
        
        The results are presented in Figure~\ref{fig:fig2}. The x-axis represents the inter-community network distance $D_{cc'}$ between two communities, and the y-axis shows the inter-community lexical distance $\Delta_{cc'}$. Each curve represents a topic: warm colours correspond to more neutral topics (sport, music, and cars), while cool colours indicate more sensitive and segregated topics (immigration, climate, and vaccines).
        In panel (a), we show the scatter plot between these two quantities for sport and immigration, obtained through bootstrap with subsampling (see Section~\ref{sec:methods} and Appendix~\ref{app:test_metrics} for details), and their moving median. In panel (b), instead, we plot the moving median and moving quartiles for all topics.
        
        In all six cases, the curves exhibit a significant positive (increasing) trend, confirmed by the Spearman's correlation coefficient $\rho$ reported at the bottom of the figure. This trend indicates that the greater the distance between communities in the retweet network, the greater the difference in their lexicon. 
        This effect represents the first empirical component of the broader phenomenon of linguistic bubbles introduced earlier, and is observed across both neutral and polarised topics.
        Indeed, even in the case of a neutral topic like sport, it is plausible that different communities discuss different sports or teams and interact less through retweets, resulting in distinct lexical communities that reflect linguistic divergences.

        To validate the observed correlation between network and lexical distances, we tested our results against a null model. Specifically, we generated 1000 randomised samples by rewiring the bipartite retweet network between influencers and users, thereby destroying the original community structure while preserving node activity (see Section~\ref{sec:methods} for details). We then repeated the full analysis on each randomised instance to compute the distribution of Spearman correlation coefficients under the null hypothesis. The comparison with real data confirms that the observed correlations are statistically significant in all cases ($p < 0.05$), as reported in Table~\ref{tab:null_hsbm}.
        
        Finally, in Appendix~\ref{app:louvain}, we replicate the same analysis using the Louvain community detection algorithm~\cite{blondel2008fast}. The consistency of the results confirms that our findings are not an artefact of the specific clustering method adopted.

        Overall, these findings indicate a meaningful relationship between structural proximity in the retweet network and lexical similarity across communities. While this analysis does not imply causation, it provides strong evidence that communication patterns reflect the linguistic makeup of discourse communities. 
        As such, it constitutes the first empirical evidence of what we have called language bubbles: communities that interact little with each other and develop increasingly divergent linguistic repertoires.

    \subsection{Linguistic diversity decreases in isolated discourse communities}
    \label{sec:results_3}

        \begin{figure*}[tbhp]
            \centering
            \includegraphics[width=0.85\linewidth,]{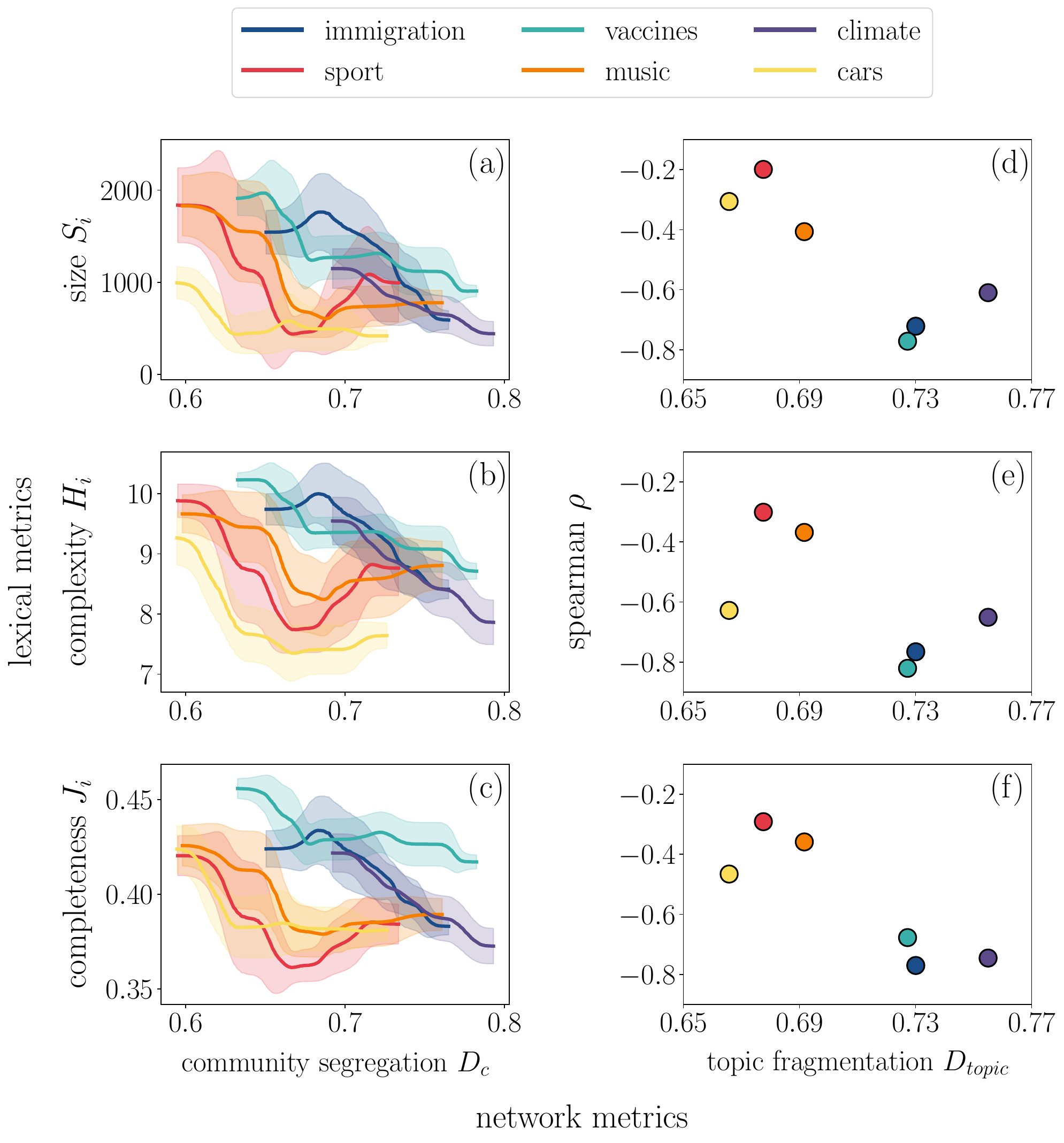}

            \caption{\textbf{Lexical scores as a function of community segregation and topic fragmentation in the retweet network.} 
            \textbf{(a)–(c)}: Lexical metrics plotted against the community network segregation $D_c$ (see Section~\ref{sec:methods}). Each curve represents a different topic, with warm colours corresponding to less segregated topics and cool colours to more segregated ones. Curves show the moving average over subsamples of user accounts, and shaded areas indicate the interquartile confidence interval. In general, more segregated communities (higher $D_c$) exhibit lower values of vocabulary size $S_i$, lexical complexity $H_i$, and completeness $J_i$, a second manifestation of the language bubbles phenomenon. This trend is stronger for more fragmented topics, with Spearman’s $\rho$ values closer to $-1$. Correlations are statistically significant, as confirmed by null model tests (Table~\ref{tab:null_hsbm}).
            \textbf{(d)–(f)}: Summary of results across topics. Each plot reports the Spearman’s $\rho$ correlation between each lexical metric and the overall topic fragmentation $D_{\text{topic}}$ (see Section~\ref{sec:methods}). Higher topic fragmentation is associated with a more pronounced reduction in lexical diversity and complexity.}
            
            \label{fig:fig3}
        \end{figure*}

        \begin{table*}[t]
        \begin{tabular}{|c|c|c|c|c|c|c|c|c|c|c|c|c|}
        \hline
         & \multicolumn{3}{c|}{\textbf{lexical distance}} & \multicolumn{3}{c|}{\textbf{vocabulary size}} & \multicolumn{3}{c|}{\textbf{complexity}} & \multicolumn{3}{c|}{\textbf{completeness}} \\
        \hline
        \textbf{topic} & $\rho_{\text{data}}$ & $\rho_0$ & $p$ & $\rho_{\text{data}}$ & $\rho_0$ & $p$ & $\rho_{\text{data}}$ & $\rho_0$ & $p$ & $\rho_{\text{data}}$ & $\rho_0$ & $p$ \\
        \hline
        immigration & 0.594 & -0.246 & 0.038 & -0.722 & 0.475 & 0.008 & -0.766 & 0.427 & 0.008 & -0.770 & 0.422 & 0.012 \\
        \hline
        vaccines & 0.580 & -0.021 & 0.009 & -0.772 & 0.487 & 0.005 & -0.821 & 0.450 & 0.002 & -0.677 & 0.338 & 0.013 \\
        \hline
        climate & 0.642 & -0.418 & 0.044 & -0.610 & 0.478 & 0.039 & -0.651 & 0.491 & 0.027 & -0.746 & 0.482 & 0.032 \\
        \hline
        sport & 0.521 & -0.357 & 0.019 & -0.199 & 0.503 & 0.023 & -0.301 & 0.469 & 0.021 & -0.291 & 0.478 & 0.024 \\
        \hline
        music & 0.314 & -0.195 & 0.005 & -0.407 & 0.362 & 0.001 & -0.368 & 0.295 & 0.009 & -0.359 & 0.329 & 0.004 \\
        \hline
        cars & 0.513 & -0.369 & <0.001 & -0.306 & 0.481 & 0.001 & -0.628 & 0.431 & <0.001 & -0.466 & 0.471 & <0.001 \\
        \hline
        \end{tabular}

        \caption{\textbf{Null model tests.} 
        Statistical validation of the results presented in Figure~\ref{fig:fig2} and Figure~\ref{fig:fig3}. As a null model, we generated 1,000 randomised configurations by rewiring the bipartite retweet network between influencers and users (before the projection). This procedure removes the underlying retweet structure, producing community assignments that are independent of actual behaviour of accounts. We then repeated the full analysis on each randomised sample, recording the Spearman’s $\rho$ coefficients between network and lexical metrics. The table reports, for each metric, the Spearman coefficient in the real data ($\rho_{\text{data}}$), the average coefficient in the null samples ($\rho_0$), and the associated $p$-value. All observed correlations are statistically significant ($p < 0.05$). Results are shown for both inter-community distance correlations (Figure~\ref{fig:fig2}) and lexical scores—vocabulary size, complexity, and completeness (Figure~\ref{fig:fig3}).}
        
        \label{tab:null_hsbm}
        \end{table*}
                
        In the previous section, we observed how discourse communities with similar audience reach—reflected by their proximity in the retweet network—also exhibit similar patterns of lexical usage. On the contrary, distant communities in the retweet network also diverge in their word choices within the same topic. 
        We now aim to assess whether the most extreme social network niches---defined as the most isolated groups in topic discussions---experience measurable costs in terms of linguistic richness and expressive complexity.

        To quantify this, we firstly define \textit{community network segregation} as the average of the inter-community network distance between a community and all others (see  Eq.~\eqref{eq:D_c} in Section~\ref{sec:methods}).
        By computing the average distance in the retweet network between a community and the other ones, we can identify the most and least segregated groups as those that are, respectively, most and least distant from all others, on average.  

        Secondly, we introduce three lexical metrics to analyse the language used within each community in relation to the overall linguistic landscape. These metrics allow us to better investigate the relationship between social segregation and the linguistic quality of discourse. Each metric captures a distinct aspect of language usage, and is first computed at the individual user level. We then aggregate the values by averaging over all users belonging to the same community.
        
        Specifically, for each user $i$, we consider the following three metrics:

        \begin{itemize}
            \item \textbf{vocabulary size} $S_i$, i.e., the number of distinct words used by the user. A larger vocabulary reflects a richer and more varied discourse, while a smaller vocabulary indicates a narrower range of expressions \cite{kohler2008quantitative, di2024evolution};
            
            \item \textbf{complexity} $H_i$, i.e., the Shannon Entropy \cite{shannon1948mathematical} of the lexical distribution of the content produced by the user. This metric quantifies the unpredictability of word choices. Higher entropy indicates a more balanced and diverse use of language, while lower values suggest that the discourse is dominated by a few recurring words, leading to a simpler language;

            \item \textbf{completeness} $J_i$, i.e., a metric that uses the Jensen-Shannon Distance \cite{cover1999elements} to compare the user's word distribution to the overall word distribution across all communities. Higher completeness (i.e., smaller distance) indicates that the user employs a broader range of the available vocabulary specific to the topic, while lower completeness implies a vocabulary which is more distinct and separated from the general discourse.
        \end{itemize}

        Section~\ref{sec:methods} reports further details about each of these metrics. These metrics provide a comprehensive picture of the diversity and complexity of language used within each community. We tested their effectiveness in capturing differences across texts of varying complexity in Appendix~\ref{app:test_metrics}.

        Figure~\ref{fig:fig3}(a-c) presents the results for the six topics. The three plots display the three lexical scores on the vertical axis, with higher scores indicating a richer, more complex, and more complete use of language. On the horizontal axis, instead, communities located further to the right are the most socially segregated. 

        The figure shows a statistically significant association between community segregation and lexical metrics, further validated against null models (results and $p$-values are reported in Table~\ref{tab:null_hsbm}). Less isolated communities tend to display greater token diversity, as indicated by their larger vocabulary size. These communities also exhibit higher complexity, measured by the Shannon entropy, reflecting a more complex and varied textual structure. Furthermore, they achieve higher completeness, shown by a lower Jensen-Shannon distance from the global token distribution. Conversely, communities with high levels of segregation exhibit reduced complexity, variability, and completeness in their language.

        These results show the second dimension of the language bubbles phenomenon: extreme discourse niches, those characterised by high levels of network segregation, are systematically associated with lower levels of lexical diversity and complexity. This pattern is particularly pronounced for more debated topics, such as climate, immigration, and vaccines, which are associated with stronger negative Spearman's $\rho$ values (i.e., closer to $-1$). More neutral topics, such as sport, music, and cars, are instead associated with weaker correlations, though the relationships remain statistically significant. 

        To quantify this association at the topic level, we compute the \textit{topic network fragmentation} $D_{\text{topic}}$ (see Eq.~\ref{eq:D_topic} in Section~\ref{sec:methods}), defined as the average network segregation $D_c$ across communities within a topic. 
        This measure formalises the visual intuition provided in Figure~\ref{fig:fig1}, allowing us to distinguish between more and less fragmented topics.
        The results across all topics are summarised in Figure~\ref{fig:fig3}(d–f), which reports the Spearman's rank correlation $\rho$, between lexical metrics and $D_c$, as a function of $D_{\text{topic}}$.
        Topics with higher $D_{\text{topic}}$, i.e., characterized by a more structurally fragmented discourse, are associated with stronger negative Spearman's $\rho$, indicating that this second dimension of the language bubbles phenomenon becomes more evident under higher levels of topic network fragmentation.
        In particular, increased structural fragmentation tends to coincide with a more substantial reduction in linguistic complexity and richness within the most segregated niches.

        As a robustness check, we replicate the same analysis using the Louvain community detection algorithm~\cite{blondel2008fast} (see Appendix~\ref{app:louvain}). The consistency of the results confirms that our findings do not depend on the specific method used for community detection.

\section{Discussion}

Online social platforms, such as X (formerly Twitter), offer a unique environment where users can connect and access content from a wide variety of sources. While these systems are, in principle, designed to facilitate exposure to diverse viewpoints, the dynamics of communication within them can lead to forms of social segregation, whereby users cluster into well-defined communities with limited interaction across group boundaries. These isolated discursive niches, documented across multiple platforms, can restrict users’ perspectives and alter the structure of public discourse.

In this work, we investigated whether this structural fragmentation is mirrored in the linguistic patterns of online communities. Our analysis of the retweet network reveals two complementary effects. First, we find that structurally closer communities---i.e., communities sharing a larger portion of their audience---also tend to exhibit more similar linguistic patterns (Figure~\ref{fig:fig2}). 
Second, and crucially, we discovered that linguistic diversity and richness systematically decline in the most segregated communities (Figure~\ref{fig:fig3}). In particular, more isolated groups tend to use vocabularies that diverge more strongly from the global lexical distribution, indicating a narrowing of expression and a weakening of shared language across the network.

These findings point to the emergence of {\em language bubbles}: environments in which language use is shaped and constrained by social proximity, leading to convergence within the community and divergence across them. In these bubbles, community members develop increasingly specialised and internally coherent linguistic repertoires while drifting away from the broader discursive norms of the platform. As Figure~\ref{fig:fig3} shows, isolated communities not only exhibit lower lexical richness and complexity but also display greater distance from the global vocabulary, highlighting the erosion of a shared linguistic ground capable of supporting cross-community communication.
    
While our results establish a robust association between social structure and linguistic behaviour, they do not resolve the question of causality. It remains unclear whether linguistic isolation is a consequence of social fragmentation or whether communities with limited lexical variability tend to form more insular social structures. Both scenarios are plausible, and further research is needed to clarify the mechanisms behind the emergence of linguistic bubbles.

Our findings offer a novel perspective on how information ecosystems may evolve towards fragmentation in structural and ideological terms and language use. However, several limitations should be acknowledged.

First, our analysis is limited to the production of a specific subset of users: political leaders and information providers active on X. This population is not representative of the general public and is likely to be influenced by distinct strategic considerations. Political and media actors often tailor their language to different audience segments using sophisticated targeting tools, which may introduce communicative asymmetries not accounted for in our model. For instance, the level of education, ideological orientation, or sociolinguistic characteristics of the target audience may shape lexical choices. In this light, our observations may also contribute to ongoing discussions on populist communication strategies, which are often characterised by simplified, emotionally resonant, and audience-specific language. Future work could further investigate this aspect by examining more refined linguistic features, such as the use of verb tenses, which might serve as indicators of rhetorical simplification or narrative framing strategies.

Second, while we observe a decline in vocabulary size and completeness in more segregated communities, our analysis does not capture the temporal dynamics of language use. Isolated communities are known to develop unique forms of expression, including the creation of neologisms and the semantic repurposing of existing terms, a phenomenon often referred to as semantic drift~\cite{liberman2012semantic}. These processes can facilitate innovation and symbolic boundary-building within communities. However, our results suggest that such innovations, if present, are not sufficient to counterbalance the overall reduction in vocabulary richness observed in more segregated groups. Specifically, the lower vocabulary size in these communities indicates that any such lexical innovations do not significantly expand the expressive range of the language. A temporal or diachronic analysis of language change would be needed to more directly assess whether the observed divergences result from active innovation or narrowing and simplification.

Finally, our analysis focuses primarily on word frequencies without considering more complex linguistic structures such as syntax, semantic context, or discourse patterns~\cite{dibona2025dynamics}. Future research could benefit from the use of semantic embedding models or transformer-based language models to explore deeper meanings and contextual relationships in language~\cite{ding2023same}. Additionally, our study is based entirely on Italian-language data, which limits the generalisability of the findings. Replicating this analysis across different linguistic and cultural contexts would help assess whether the emergence of language bubbles is a universal feature of fragmented online discourse or a context-dependent phenomenon.

\section{Methods}
\label{sec:methods}

    \subsection{Data Preprocessing}

    In this section, we describe the process used to collect and preprocess the data for our analysis.

        \subsubsection{Dataset creation}

            Using the Twitter/X API, we collected tweets from a total of $583$ monitored accounts spanning January 2018 to December 2022, prior to the restrictions introduced by the platform’s new management\footnote{https://twitter.com/XDevelopers/status/1621026986784337922}. Additionally, we included retweets of the most engaging content, specifically those with at least 20 retweets.
        
            The raw data comprises approximately $14$ million tweets collected by tracking the activity of the monitored accounts. Of these accounts, $546$ profiles represent key players in Italian online discourse (e.g., \textit{La Repubblica}, \textit{Il Corriere della Sera}, \textit{Il Giornale}). The profiles were selected based on the list of news sites monitored by NewsGuard, a news rating agency that evaluate the reliability of outlets. The list is estimated to account for $95\%$ of online news engagement, ensuring nearly comprehensive coverage of news-related social dialogue~\cite{ne2022}. To complement this, we additionally included $37$ profiles of Italian political entities~\cite{brugnoli2024comm}, encompassing all major political parties and their leaders (e.g., \textit{Giorgia Meloni} and \textit{Brothers of Italy}, \textit{Elly Schlein} and \textit{Democratic Party}, \textit{Giuseppe Conte} and \textit{Five Stars Movement}).
                        
            From this corpus, we extracted datasets focused on six different topics: three more socially and politically sensitive ones (immigration, climate, and vaccines) and three more neutral ones (cars, sport, and music).
            Each dataset was constructed by defining a set of keywords or root forms of relevant terms in the topic, and selecting all tweets containing these keywords. The specific keywords for each topic are reported in Sec.~\ref{app:dataset_details}. The number of accounts with at least one selected tweet ranges between $320$ and $518$ per topic, with the number of tweets in each dataset on the order of $10^5$. Detailed statistics on the selected accounts and their tweets are also provided in the Appendix~\ref{app:dataset_details}.
        
        \subsubsection{Retweet network} 
            
            To assess segregation in the system, we track user activity based on retweets. First, we process the retweet data to identify the most active users. We construct a bipartite graph for each topic, with one layer representing users who retweet, which we refer to generally as \textit{accounts}, and the other representing users who produced tweets and are retweeted, which we refer to as \textit{influencers}. A link is formed when an account in the first layer retweets an influencer from the second layer, resulting in a weighted bipartite graph where the biadjacency matrix $\boldsymbol{W}_{ij}$ counts how many times user $i$ retweeted user $j$. The first layer contains $N_1$ accounts, the second $N_2$ influencers, with a total number of connections equal to $E$. We report the details for each topic in Appendix~\ref{app:dataset_details}.

            We focus on the layer of influencers and project the bipartite retweet network onto it, resulting in a weighted monopartite network of influencer co-occurrence. This results in a weighted network, where the adjacency matrix $w_{ij}$ is given by:
            \begin{equation}\label{eq:W}
                w_{ij} = \dfrac{\sum_k W_{ik} W_{kj}}{ \sqrt{\sum_k W_{ik}^2} \sqrt{\sum_k W_{kj}^2} }
            \end{equation}
            Here, $w_{ij}$ measures the co-occurrences between influencer $i$ and $j$, based on how often both are retweeted by the same set of accounts. This value acts as a similarity measure between influencers. 
            
            After projection, we filter out isolated nodes that have no co-occurrence links with any other account, retaining only the largest connected component of the resulting graph. The number of influencer nodes in the final monopartite network ranges from 96 to 135 accounts, depending on the topic. We then apply a pruning procedure~\cite{zhou2012} to remove weaker links and retain only the most significant connections, ensuring that the network's connected component reflects meaningful relationships.

            Figure~\ref{fig:fig1} illustrates two examples of the resulting retweet networks for the topics \textit{immigration} and \textit{sport}, as obtained through this procedure.

        \subsubsection{User lexical network}
            
            For each influencer, we compute the empirical distribution of word frequencies based on their tweets. We first clean and tokenise the tweet content using the \texttt{ntlk} Python module~\cite{Loper2002}. Stopwords are removed with the \texttt{stopwords} module from \texttt{nltk.corpus}, and the content is lemmatised using \texttt{WordNetLemmatizer}. After preprocessing, we construct a user-token weighted bipartite network, where each influencer is connected to the tokens they used, with multiplicity. The token layer consists of between $10^4$ and $10^5$ nodes, depending on the topic. The adjacency matrix $\boldsymbol{M}$ of the resulting network describes how many times user $i$ employed token $\alpha$ through the value $M_{i \alpha}$.

            We then filter this network to retain only connections that reflect distinctive vocabulary use by each user. To do so, we exploit the concept of the \textit{kernel lexicon}~\cite{cancho2001small, bellina2025cognitive}, which refers to the set of words that are commonly used by all speakers and do not capture individual or topical differences. These high-frequency words contribute noise by obscuring meaningful linguistic variation. Since our dataset is entirely in Italian, the kernel lexicon is largely uniform across all topics. Therefore, we approximate and remove it by excluding the 1,000 most frequent tokens in the corpus. Further details on the kernel lexicon, including its estimated size and frequency threshold, are provided in Appendix~\ref{app:kernel}.
            
            Finally, we compare the user-token bipartite network with a proper statistical benchmark, to retain only significant connections. In the present case, we used the BiWCM (\emph{Bipartite Weighted Configuration Model}~\cite{bruno2023inferring}). In a nutshell, BiWCM is a maximum entropy null model, informed only about the strength sequence, i.e., the total number of tokens each user used and the number of times a token was used. Being maximally entropic, the null model is unbiased by definition (see Ref.~\cite{Cimini2019} for a general reference) and therefore can uncover non-trivial local structures.

            After removing the kernel lexicon and filtering out statistically insignificant entries in $M_{i\alpha}$, we obtain a validated bipartite network that retains only meaningful user–token associations, allowing us to reconstruct the token probability distribution for each influencer. Normalising each row of the adjacency matrix we obtain the relative frequency distribution of tokens $\alpha$ used by each user $i$, which we call \textit{rating} \cite{bellina2023effect}:
            \begin{equation}
                r_{i \alpha} = \dfrac{M_{i \alpha}}{\sum_{\beta} M_{i \beta}}.
                \label{eq:ratings}
            \end{equation}
            We interpret these values as the validated empirical probability distributions of token usage for each user. They form the basis for analysing the lexical properties of both individual influencers and their communities, as discussed in Sections~\ref{sec:results_2} and~\ref{sec:results_3}.

    \subsection{Data Analysis}

            In this section, we describe the technical procedures used to analyse the data, with a focus on measuring network segregation and assessing the lexical diversity and complexity of user-generated content.

        \subsubsection{Clustering methods}
            We detect communities within the monopartite retweet network using the hierarchical stochastic block model (hSBM)~\cite{peixoto2014hierarchical,peixoto2017nonparametric}. This method has been developed to find statistically significant clusters at multiple hierarchical levels for the analysis of generic weighted networks, without obtaining spurious modular structures due to noise or randomness.
            As the name suggests, the model produces a hierarchical clustering, providing a richer structure of the data encoded in the adjacency matrix $w_{ij}$.
            
            To ensure the robustness of our results, we performed 100 iterations of the algorithm. 
            Notice that the number of clusters and levels of granularity obtained is not fixed, but is automatically suggested by the algorithm.
            By running the algorithm multiple times, we aimed to capture the inherent variability and uncertainty in the Monte Carlo partitioning process. 
            Subsequently, the consensus partition is calculated by maximising the overlap among all partitions obtained in the 100 runs.
            Such a consensus partition serves as a robust representation of the underlying structure within the analysed data.

            Once we have obtained the consensus hierarchical partition with multiple levels, we have removed the first most granular level, since almost all modules at this level contained only one node. We hence considered the second level, and computed its minimum description length to use as a baseline for the model selection criterion~\cite{peixoto2015model}. At this level, there were still between zero and four modules with fewer than 10 nodes, which introduced noise into the analysis. To address this, we merged each small module with another module from the same branch at a higher level of the hierarchy, choosing the merge that minimised the overall description length. This procedure was applied to all modules with fewer than 10 nodes, and the search was extended to higher-level branches if necessary.

            The community partitions obtained with this method were used in the analyses presented in Sections~\ref{sec:results_2} and~\ref{sec:results_3}, where we examined the relationship between lexical metrics and network-based measures. The number of users in each community varies by topic, and detailed statistics are provided in Appendix~\ref{app:dataset_details}.
            
            For comparison, we also applied the Louvain algorithm~\cite{blondel2008fast} to the same data, detecting clusters by maximising network modularity. In cases where the algorithm produced clusters with fewer than 10 nodes, these were merged with the closest cluster, based on the distance metric defined in Eq.~\ref{eq:D_cc'}.
            
            In Section~\ref{app:louvain}, we replicate the main analyses using Louvain-based community partitions to assess the robustness of our results.
            
        \subsubsection{Network metrics}

            In our analysis we compute various measures on the retweet network, where edge weights $w_{ij} \in [0,1]$ represent the similarity (co-occurrence) between nodes. To define a distance metric from these similarity values, we transform the weights as $\overline{w}_{ij} = 1 + \lambda - w_{ij}$, where $\lambda > 0$ is a regularization parameter introduced for normalisation purposes. This transformation is motivated by the fact that we use inverse distances in the following analysis. If we set $\lambda = 0$, the resulting distances $\overline{w}_{ij}$ would lie in the interval $[0,1]$, and their inverses could become arbitrarily large as $w_{ij} \to 1$. By introducing a positive $\lambda$, we ensure that distances are bounded in $[\lambda, 1 + \lambda]$, which in turn guarantees that their inverses are also bounded and numerically stable. Moreover, this formulation allows for network distances to be normalised between 0 and 1 after inversion, making them comparable across different networks or settings. In all our analyses, we fix $\lambda = 1$.

            Based on these pairwise distances $\overline{w}_{ij}$, we compute the weighted shortest path length $d_{ij}$ between each pair of influencers $i$ and $j$. $d_{ij}$ coincides with $\overline{w}_{ij}$ if $i$ and $j$ are directly connected. Otherwise, if $i$ and $j$ are not nearest neighbours, $d_{ij} = min_{\{i, k_1, k_2, ..., k_n, j\}} (\overline{w}_{ik_1}+\overline{w}_{k_1k_2} +...+\overline{w}_{k_{n-1}k_n}+\overline{w}_{k_{n}j})$, where $\{i, k_1, k_2, ..., k_n, j\}$, $\forall n\ge 1$, represents all the paths connecting $i$ and $j$ through intermediate nodes. Using this shortest path length, we measure the average separation between communities. The inter-community distance between two communities $c$ and $c'$ (as used in Section~\ref{sec:results_2}) is defined as:
            \begin{equation}
                D_{cc'} = 1 - \lambda \dfrac{1}{n_c n_{c'}} \sum_{i \in c, j \in c'} \dfrac{1}{d_{ij}}
                \label{eq:D_cc'}
            \end{equation}
            Here, $n_c$ and $n_{c'}$ are the number of influencers in communities $c$ and $c'$, respectively. The measure is based on the average inverse shortest-path distance between all pairs of nodes across the two communities. Using inverse distances ensures that disconnected node pairs contribute 0 rather than diverging, since their path length is undefined (infinite). Finally, reintroducing the parameter $\lambda$ guarantees that $D_{cc'}$ remains bounded within $[0,1]$.

            This metric quantifies the extent to which users from the two communities are retweeted by the same accounts. If no user retweets from both communities, $D_{cc'} = 1$, indicating maximal distance between them. Conversely, smaller values indicate that the two communities are frequently retweeted by the same accounts, reaching $D_{cc'} = 0$ in the extreme cases where the two communities share exactly the same audience. This metric is employed for the analysis of Section~\ref{sec:results_2}.
            
            We then assess the average level of segregation for a community, reflecting its degree of separation from all the others. The community network segregation of $c$ is defined as:
            \begin{equation}
                D_c = \dfrac{1}{N_c - 1} \sum_{c' \neq c} D_{cc'} 
                \label{eq:D_c}
            \end{equation}
            where $N_c$ is the total number of communities within a topic. This metric, for community $c$, is essentially the inter-community network distance $D_{cc'}$ averaged across all other communities $c'$. The presence of $\lambda$ is again needed for bounding $D_c \in [0,1]$. A higher value of community network segregation indicates a greater degree of isolation from the rest of the network. In particular, in a maximally segregated scenario where there are no interconnections between $c$ and the other communities, we have $D_c = 1$. In a minimally segregated scenario, where all interconnections among communities exist and are maximal $w_{ij} = 1$, then $D_c = 0$. We use this metric in the analysis presented in Section~\ref{sec:results_3}, where we investigate the relationship between network segregation and lexical scores.

            Finally, we quantify the overall level of network fragmentation within a topic by averaging the community segregation scores:
            \begin{equation}
                D_{topic} = \dfrac{1}{N_c} \sum_{c=1}^{N_c} D_c
                \label{eq:D_topic}
            \end{equation}
            This score captures the average structural isolation among communities discussing a given topic, with values of $D_{topic}$ closer to 1 indicating strong fragmentation, whilst the opposite is observed for $D_{topic} \approx 0$.
            Different topics exhibit varying levels of fragmentation, as illustrated in Figure~\ref{fig:fig1}, where the contrasting values of $D_c$ between two example topics can be visually assessed.

        \subsubsection{Lexical metrics}

            We explore lexical differences between communities by computing distances between token distributions of their users, using the Jensen-Shannon divergence \cite{cover1999elements}. The lexical distance between two users $i$ and $j$, each with their token distribution (ratings, as defined in Eq \ref{eq:ratings}) $\vec{r}_i = \{r_{i \alpha} \}$ and $\vec{r}_j = \{r_{j \alpha} \}$, is computed as the Jensen-Shannon divergence between their token distributions:
            \begin{equation}
                \delta_{ij} = J(\vec{r}_i \ || \ \vec{r}_j) = H(\vec{R}) - \dfrac{1}{2}[H(\vec{r}_i) + H(\vec{r}_j)]
                \label{eq:delta_ij}
            \end{equation}
            where $\vec{R}_{ij} = (\vec{r}_i + \vec{r}_j) / 2$ is the combined distribution, and $H(\cdot)$ is the Shannon Entropy of the token distributions:
            \[
                H(\vec{r_i}) = - \sum_{\alpha} r_{i \alpha} \log r_{i \alpha}
            \]
            The metric $\delta_{ij}$ measures the divergence between the probability distributions of two users, with a value of 0 indicating identical distributions and higher values reflecting greater dissimilarity.

            To compare entire communities, we aggregate lexical distances across all user pairs. The lexical distance between two communities $c$ and $c'$ is defined as the average lexical distance between all users belonging to those communities:
            \begin{equation}
                \Delta_{cc'} = \dfrac{1}{n_c n_{c'}} \sum_{i \in c, j \in c'} \delta_{ij}
                \label{eq:Delta_cc'}
            \end{equation}
            This metric allows us to quantify lexical divergence both between individual users and across entire communities. It is used in Section~\ref{sec:results_2} to examine the relationship between lexical and structural distances.
            
            We then employed several metrics to characterise the content of influencers' tweets, focusing on quantifying the diversity, complexity, and richness of the language used by different communities.

            First, we evaluate the \textit{vocabulary size} $S_i$ for each user, which simply counts the number of distinct tokens used by a user: 
            \begin{equation}
                S_i = | \{ r_{i \alpha} \neq 0 \} |
            \end{equation}
            A higher number of unique tokens reflects greater linguistic diversity, while a lower number suggests a limited range of expression. Note that the vocabulary size is computed after applying both the kernel lexicon removal and the statistical validation procedures described in the previous sections.

            Secondly, we assess \textit{complexity} using Shannon Entropy, which measures the unpredictability of a user's word choices. This metric captures how balanced and complex the language used is. The lexical complexity $H_i$ of user $i$ is given by the Shannon Entropy of its ratings distribution:
            \begin{equation}
                H_{i} = H(\vec{r}_i) =  - \sum_{\alpha} r_{i \alpha} \log r_{i \alpha}
            \end{equation}
            where $r_{i \alpha}$ denotes the normalised frequency of token $\alpha$ used by user $i$, as defined in Eq.~\ref{eq:ratings}.
            Higher values of $H_i$ indicate greater lexical unpredictability and a more complex distribution of token usage, reflecting a higher amount of information contained in the user's language. Conversely, lower values of $H_i$ reflect more predictable and repetitive language, where a small set of tokens dominates the user's discourse, resulting in lower complexity of the content.
            
            Finally, \textit{completeness} is measured using the Jensen-Shannon Distance between a user’s token distribution and the overall token distribution across all users. A higher level of completeness (i.e., greater similarity to the overall token distribution) indicates that the user employs a broad and information-rich vocabulary, covering a wide range of topics and lexical contexts. Conversely, lower completeness reflects a narrower and more specialised use of language, suggesting a degree of lexical isolation from the general discourse. The lexical completeness of user $i$ is calculated as:
            \begin{equation}
                J_i = 1 - J(\vec{r}_i \ || \ \vec{Q}) = 1 - 
                H(\vec{R}_{i}) + \dfrac{1}{2}[H(\vec{r}_i) + H(\vec{Q})]
            \end{equation}
            where $\vec{r}_i$ is the token distribution of user $i$, while $\vec{Q}$ is the overall token distribution aggregated over all users.
            Moreover, $\vec{R}_i = (\vec{r}_i + \vec{Q}) / 2$ is the usual combined distribution, needed for the computation of the Jensen-Shannon divergence. On the one hand, a value of $1$ indicates identical distributions, when the user's language spans the entire global distribution of tokens. On the other hand, a value closer to $0$ reflects greater divergence, meaning that the user's language is more limited compared to the general discourse.

            These user-level metrics are then aggregated at the community level by averaging over all users within each community. Letting $q_i$ denote one of the three lexical scores for user $i$, the corresponding score for community $c$ is defined as:
            \[
                q_c = \dfrac{1}{n_c} \sum_{i \in c} q_i
            \]
            where $n_c$ is the number of users in community $c$. These community-level scores are used in the analysis presented in Section~\ref{sec:results_3} to examine the relationship between lexical properties and network segregation.

            The effectiveness of the three proposed lexical metrics (vocabulary size, complexity and completeness) in distinguishing between texts of varying complexity is evaluated in Appendix~\ref{app:test_metrics}.
            
            \subsubsection{Subsampling procedure}
            
            Figures~\ref{fig:fig2} and~\ref{fig:fig3} are produced using a subsampling procedure designed to test the robustness of our findings against incomplete coverage of the dataset. Since our topic-based extraction does not guarantee comprehensive representation of the discourse, we adopt a repeated random sampling strategy to mitigate potential biases due to missing or uneven data.
            
            Metrics for the subsampled communities are computed on the original lexical and network properties of users, preserving their pre-subsampling values. In particular, we avoid artificially inflating network distances, since they depend on shortest paths, and our approach thus provides conservative estimates.
            
            In Figure~\ref{fig:fig2}, for each community, we generate 1,000 random subsamples, each including 60\% of its members. For each subsample, we compute the lexical distances between this reduced group and the full (non-subsampled) other communities. Each point in the figure represents one such inter-community distance, enabling us to build distributions while preserving the internal variability of each group.
            
            In Figure~\ref{fig:fig3}, the same 60\% subsampling is applied independently to each community. For each draw, we compute the average lexical scores (vocabulary size, complexity, completeness) across the sampled users. Each point thus reflects the aggregate lexical profile of a subsampled community, with the procedure repeated 1,000 times.
            
            This approach allows us to estimate confidence intervals and control for noise due to unbalanced or sparse data. However, it also introduces two potential limitations. First, the arbitrary choice of the 60\% threshold may influence the stability of the results: smaller percentages would increase variance, while larger ones might reduce the benefit of resampling. Second, subsampling may not fully capture highly skewed distributions within communities, for example, if a few highly active users dominate the lexical signal.

        \subsubsection{Null models}

            Since we aim to assess whether our results are merely an artefact of the partitioning of users into communities, we introduce a null model in which communities are randomly rearranged. This is achieved by randomising the edges of the bipartite retweet network $W_{ij}$ between accounts and influencers, and then recomputing all analyses, both on the retweet network structure and on the lexical properties.
            
            Through this process, the projected retweet network of influencers $w_{ij}$ no longer contains any structural information about the discursive communities present in the dataset. Consequently, we effectively discount the impact of the community partitioning procedure. 
            
            We then regenerate the plots shown in Figures~\ref{fig:fig2} and~\ref{fig:fig3}, computing the Spearman's correlation coefficients $\rho$ and assessing the significance of the correlations found in the original (real data) case. The p-value is then determined from the distribution of $\rho$ values obtained by randomising the retweet network 1,000 times, counting the fraction of times the random Spearman $\rho$ was higher than the actual one. This allows us to quantify the statistical significance of the correlations observed in the real dataset.    

            The results of the validation tests are reported in Table~\ref{tab:null_hsbm} for the hSBM clustering method, and in Table~\ref{tab:null_louvain} in Appendix~\ref{app:louvain} for the Louvain-based partitioning.

\section*{Acknowledgements}

This work has been partially supported by the Horizon Europe VALAWAI project (grant agreement number 101070930).
G.D.B. acknowledges support of the French Agence Nationale de la Recherche (ANR), under grant  ANR-21-CE38-0020 (project ScientIA). 
F.S. was partially supported by the project ``CODE – Coupling Opinion Dynamics with Epidemics'', funded under PNRR Mission 4 ``Education and Research'' - Component C2 - Investment 1.1 - Next Generation EU ``Fund for National Research Program and Projects of Significant National Interest'' PRIN 2022 PNRR, grant code P2022AKRZ9.

\section*{Author contributions statement}

D.R.L.S. conceptualized the original idea. A.B., R.L.S., V.L., and G.D.B. designed the study. D.R.L.S. and E.B. collected and curated the data and conducted preliminary analyses. F.S. and P.G. supervised the methodological framework. A.B. and G.D.B. carried out the experiments and performed the data analysis, and compiled an early draft of the manuscript. G.D.B. coordinated the overall research activities. V.L. provided scientific supervision, guidance throughout the project and funding acquisition. All authors contributed to data interpretation, discussed the results, reviewed the draft, and approved the final version of the manuscript.

\section*{Additional information} 

All code used in the analysis of this paper is available at the repository: \url{https://github.com/alebellina412/language_bubbles}.
Data are made available under the platforms’ terms of service. 
The authors declare no competing interests.

\clearpage
\newpage


\setcounter{figure}{0}
\setcounter{table}{0}
\setcounter{equation}{0}
\setcounter{section}{0}
\makeatletter
\renewcommand{\thefigure}{S\arabic{figure}}
\renewcommand{\theequation}{S\arabic{equation}}
\renewcommand{\thetable}{S\arabic{table}}

\renewcommand{\thesection}{S\arabic{section}}

\setcounter{secnumdepth}{2} 
\onecolumn

\begin{center}
\textbf{\Large Supplementary Information for\\``Language bubbles in online social networks"}
\end{center}

\section{Dataset details}
\label{app:dataset_details} 

    In this section, we provide quantitative details about the Twitter/X datasets used in our analysis. The raw dataset includes approximately $14$ million tweets ($13,996,467$) from $583$ users representing Italian information leaders, along with a total of $30$ million retweets ($29,131,371$) collected between 2018 and 2022.
    
    The raw dataset is filtered to create specific sub-datasets for each topic. We define a set of keywords and roots, and select all tweets containing at least one of these keywords. The keywords used for extracting the four topics are:
    
    \begin{itemize}
        \item \textbf{immigration:} \textit{immigrat-, immigrazion-, migrant-, stranier-, profug-, ong, ngo}
        \item \textbf{vaccines:} \textit{vaccin-}
        \item \textbf{climate:} \textit{clim-, riscaldamento globale, emission-, rinnovabil-, inquinament-, sostenibil-, anidride carbon-, deforest-}
        \item \textbf{sports:} \textit{sport, calcio, tennis, volley, pallavol-, basket, pallacanestr-}
        \item \textbf{music:} \textit{music-, canzon-, album, cantant-, concert-}        
        \item \textbf{cars:} \textit{automobil-, motor-, formula 1, ferrari, lamborghini, porsche}
    \end{itemize}
    
    This procedure results in four datasets, with specific characteristics reported in Table \ref{tab:a1}.

    \begin{table*}[ht!]
    \centering
    \begin{tabular}{|c|c|r|}
    \hline
    \textbf{topic} & $N_{\text{users}}$ & $N_{\text{tweets}}$ \\ \hline
    immigration & 518 & $175,807$  \\ \hline
    vaccines    & 483 & $227,788$  \\ \hline
    climate    & 327 & $88,801$  \\ \hline
    sports      & 324 & $290,607$  \\ \hline
    music      & 327 & $154,419$  \\ \hline
    cars      & 320 & $69,376$  \\ \hline
    \end{tabular}
    \caption{\textbf{Specifics of topic datasets}. We report the number of users that produce tweets $N_{\text{users}}$, the total number of tweets $N_{\text{tweets}}$.
    }
    \label{tab:a1}
    \end{table*}
    
    We then construct a bipartite network where one layer consists of users that have received retweets (influencers), and the other layer consists of users who have retweeted (accounts). We project this onto the active accounts layer to obtain a monopartite network, whose connected component may be smaller than the initial layer. Finally, we apply a clustering algorithm (hSBM or Louvain) algorithm to identify communities. The specifics for each topic are reported in Table \ref{tab:a2}.

    \begin{table}[ht!]
    \centering
    \begin{tabular}{|c|c|r|r|c|c|c|c|}
    \hline
    \textbf{topic} & $N_{\text{influencers}}$ & $N_{\text{accounts}}$ & $E_{\text{retweet}}$ & $N_{\text{mono}}$ & $E_{\text{mono}}$ & $N_{\text{comm}}$ (hSBM) & $N_{\text{comm}}$ (Louv.)\\ \hline
    immigration & 135    & $145,614$   & $1,682,826$   & 131   & $1,760$  & 5 & 4   \\ \hline
    vaccines     & 134    & $99,531$   & $752,400$   & 132    & $1,533$  & 4 & 3   \\ \hline
    climate     & 127    & $54,453$   & $191,108$   & 124    & $1,113$ & 6 & 5  \\ \hline
    sports       & 121    & $59,098$   & $126,612$   & 115    & $2,353$  & 7 & 3   \\ \hline
    music       & 140    & $81,003$   & $223,585$   & 135    & $2,765$  & 8 & 5   \\ \hline
    cars       & 100    & $25,187$   & $46,829$   & 96    & $1,643$  & 5 & 3   \\ \hline
    \end{tabular}
    \caption{\textbf{Specifics of retweet networks}. We report the number of users in the first layer of the bipartite retweet network $N_{\text{influencers}}$; the number of users in the second layer $N_{\text{accounts}}$; the number of edges $E_{\text{retweets}}$; the number of nodes in the largest connected component after the projection $N_{\text{mono}}$; the number of edges in the monopartite network $E_{\text{mono}}$; and the number of communities $N_{\text{comm}}$ identified with the hSBM algorithms, and with Louvain algorithm.}
    \label{tab:a2}
    \end{table}
    
    Afterwards, we construct a bipartite network of accounts and tokens based on the content of the tweets, after lemmatisation and removal of stopwords. Specifics for each topic are reported in Table \ref{tab:a3}. We apply the \texttt{BiWCM} algorithm to validate its entries and build the token distribution for each account.

    \begin{table}[ht!]
    \centering
    \begin{tabular}{|c|c|c|r|}
    \hline
    \textbf{topic} & $N_{\text{users}}$ & $N_{\text{tokens}}$ & $E_{\text{tokens}}$  \\ \hline
    immigration & 132 & $57,209$ & $322,153$  \\ \hline
    vaccines    & 132 & $47,962$ & 2$87,638$  \\ \hline
    climate    & 124 & $42,640$ & $193,015$  \\ \hline
    sports      & 115 & $57,849$ & $238,837$  \\ \hline
    music      & 135 & $55,282$ & $233,911$  \\ \hline
    cars      & 96 & $27,364$ & $93,706$  \\ \hline
    \end{tabular}
    \caption{\textbf{Specifics of user-token networks.} We report the number of users in the first layer of the bipartite user-tokens network $N_{\text{users}}$; the number of tokens in the second layer $N_{\text{tokens}}$; and the number of edges $E_{\text{tokens}}$.}
    \label{tab:a3}
    \end{table}

\section{Identification and removal of the kernel lexicon}
\label{app:kernel}

    \begin{figure*}[tbhp!]
        \centering
        \includegraphics[width=0.7\linewidth,]{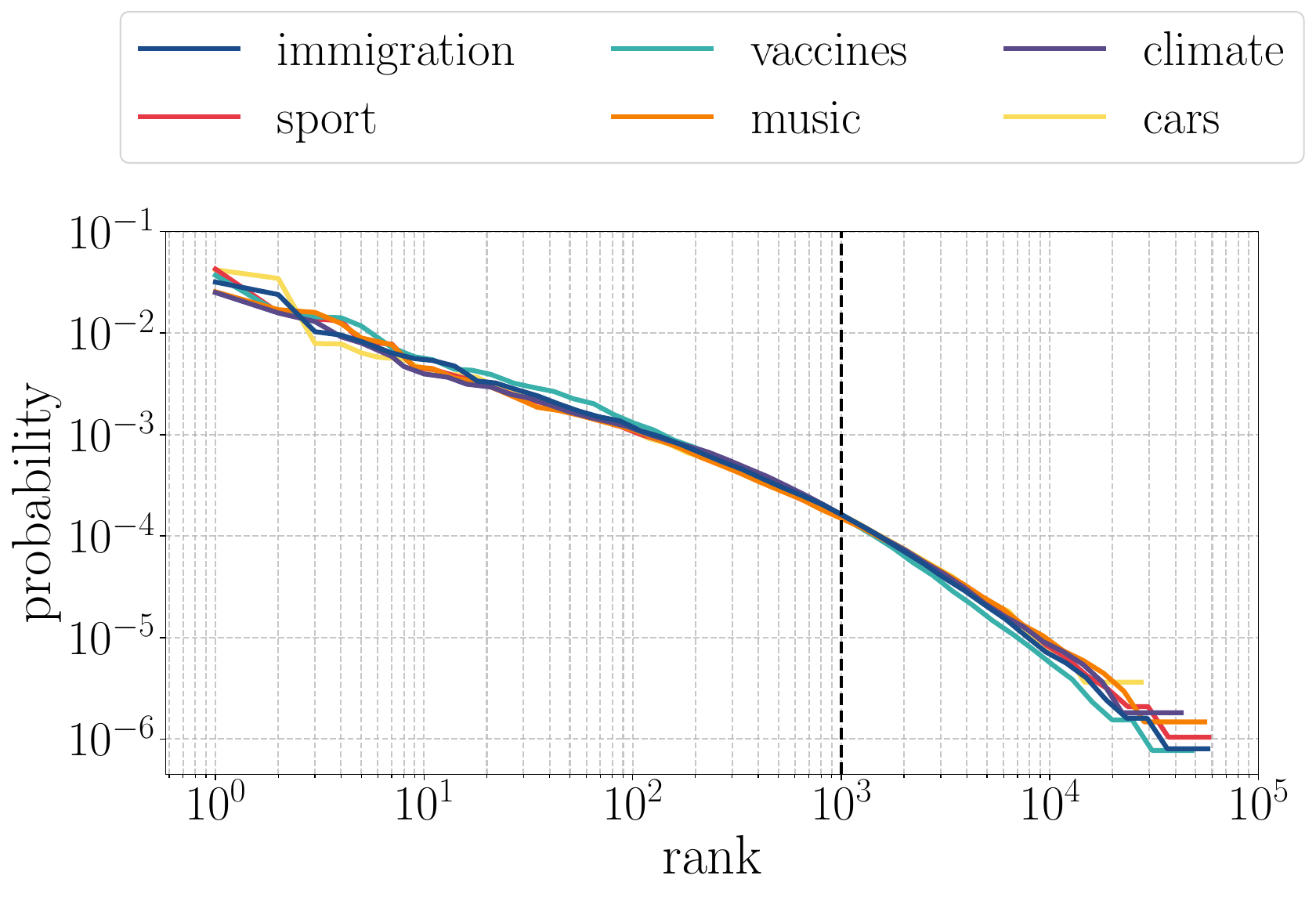}
        \caption{\textbf{Identification of the kernel and topical lexicon.} Rank size distribution of tokens (after lemmatisation and removal of stopwords) for the six topics: immigration, vaccines, climate, sport, music, and cars. The kernel lexicon, which corresponds to the first part of the distribution following $f(R) \approx R^{-1}$, consists of common and mostly uninformative words. The topical lexicon represents the second part of the distribution, following $f(R) \approx R^{-2}$. To focus solely on the topical lexicon, we identify the kernel lexicon as approximately $10^3$ words in all cases, and remove it from the distribution. The rank $R = 10^3$ is highlighted with a vertical dashed line.}
        \label{fig:rank_size}
    \end{figure*}

    Several studies have pointed out that the frequency rank distribution of words in natural language follows two distinct regimes~\cite{montemurro2001beyond, li2010fitting, bellina2025cognitive}. The most frequent words, those with smaller rank, follow a power law distribution $f(R) \sim R^{-1}$, a famous property known as Zipf's Law~\cite{zipf2013psycho}. For higher ranks, large deviations appear, and the exponent of the power law shifts from $1$ to $2$, giving $f(R) \sim R^{-2}$~\cite{ferrer2001two}. The rank-frequency distributions for the six topics, obtained after lemmatisation and stopword removal, are shown in Figure~\ref{fig:rank_size}.
    
    The first part of the distribution contains the most common words, which are used and understood by all speakers. This part, referred to as \textit{kernel lexicon}~\cite{cancho2001small}, does not really contribute to communication. The second part of the distribution, containing all those rare and specific words, has been referred to as the one containing the \textit{topicality} of the language, that is, all these particular words used by one speaker and not by another based on topics and personal ideas~\cite{gerlach2014scaling}.
    
    To extract more relevant information about the characteristics of speech produced by accounts via tweets, we want to focus only on the topical part of the distribution. We then identify the size of the kernel lexicon, which in all cases contains around $10^3$ words, and filter it out. The size of the kernel lexicon was estimated from the rank-frequency distribution (Figure~\ref{fig:rank_size}) by identifying the transition point between the two power-law regimes: from an exponent of $-1$ (Zipf's Law) to an exponent of $-2$~\cite{bellina2025cognitive}. In our case, the kernel lexicon is smaller than in full natural language, due to the thematic constraints of the dataset and, most importantly, the lemmatisation and stopword removal procedures.
    
    This is a crucial step in our analysis since the kernel lexicon, containing the most frequent words, produces large noise in the token distribution. Notice that the number of tokens is always around $5 \cdot 10^4$, so we are filtering out only about $2\%$ of the total number of tokens.

\section{Testing lexical metrics on texts of varied complexity}
\label{app:test_metrics} 

\begin{table}[ht!]
\centering
\begin{minipage}{\linewidth}

    \centering
    \includegraphics[width=0.9\linewidth]{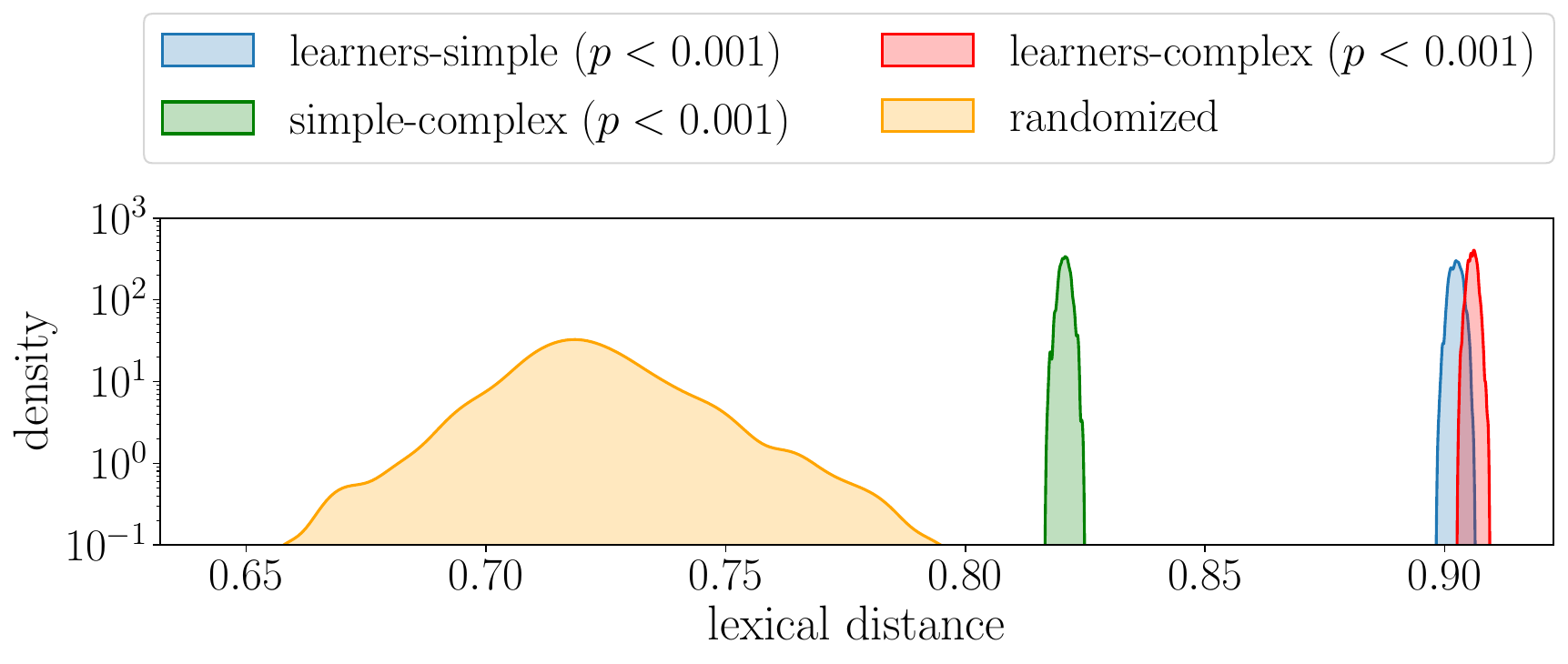}
    \captionsetup{type=figure}
    \caption{\textbf{Test of lexical distance on texts of varying complexity.} 
    Probability distributions of lexical distances computed between synthetic communities, each associated with a different text corpus of increasing complexity (learners, simple, complex). The null distribution is obtained by building synthetic communities through random sampling of users across all corpora, thereby mixing different levels of language complexity. All three empirical distributions (learners–simple, simple–complex, and learners–complex) are significantly shifted with respect to the null ($p<0.001$ in all three cases), confirming that lexical distance effectively captures differences in language complexity. The effect is strongest between learners and the other two corpora, while the distance between simple and complex texts, both representative of the native language, is smaller but still significant. Distributions are built through the usual subsampling procedure described in Section \ref{sec:methods}.}
    \label{fig:test_lexical_distances}
    \vspace{0.4cm}

    \begin{tabular}{|l|c|c|c|}
    \hline
    \textbf{type} & \textbf{vocabulary size ($S$)} & \textbf{complexity ($H$)} & \textbf{completeness ($J$)} \\
    \hline
    \textbf{learners} & 1048 $\pm$ 30 & 9.76 $\pm$ 0.04 & 0.315 $\pm$ 0.005 \\
    \textbf{simple} & 1774 $\pm$ 35 & 10.54 $\pm$ 0.03 & 0.459 $\pm$ 0.003 \\
    \textbf{complex} & 3931 $\pm$ 48 & 11.69 $\pm$ 0.02 & 0.629 $\pm$ 0.004 \\
    \hline
    \end{tabular}

    \vspace{0.4cm}

    \begin{tabular}{|l|c|c|c|c|c|c|}
    \hline
    \textbf{type} & \textbf{$\Delta S$} ($\Delta S_0 = 396 \pm 300$) & \textbf{$p_S$} & \textbf{$\Delta H$} ($\Delta H_0 = 0.26 \pm 0.19$) & \textbf{$p_H$} & \textbf{$\Delta J$} ($\Delta J_0 = 0.04 \pm 0.03$) & \textbf{$p_J$} \\
    \hline
    \textbf{learners–simple} & 727 $\pm$ 46 & 0.159 & 0.77 $\pm$ 0.05 & 0.017 & 0.14 $\pm$ 0.01 & 0.005 \\
    \textbf{simple–complex} & 2157 $\pm$ 60 & <0.001 & 1.16 $\pm$ 0.04 & <0.001 & 0.17 $\pm$ 0.01 & <0.001 \\
    \textbf{learners–complex} & 2884 $\pm$ 57 & <0.001 & 1.93 $\pm$ 0.05 & <0.001 & 0.31 $\pm$ 0.01 & <0.001 \\
    \hline
    \end{tabular}

    \vspace{0.4cm}
    \captionsetup{type=table}

    \caption{\textbf{Test of lexical metrics on texts of varying complexity}: \textbf{Top}: average values of lexical scores ($S$, $H$, and $J$) for the synthetic communities. As expected, the complex community exhibits the highest scores, learners the lowest, and simple users lie in between. \textbf{Bottom}: differences in lexical scores ($\Delta S$, $\Delta H$, and $\Delta J$) between communities, compared to null distributions of differences ($\Delta S_0$, $\Delta H_0$, $\Delta J_0$) generated by random sampling of users. $p$-values indicate the significance of observed differences against the null model. Differences involving the complex group are highly significant across all metrics. For the learners--simple comparison, vocabulary size is less effective, while complexity and completeness still show significant distinctions. These results confirm the ability of the proposed lexical scores to capture differences across language samples of varying complexity.}

    \label{tab:test_lexical_scores}
\end{minipage}
\end{table}

    In this appendix, we evaluate the lexical metrics introduced in Section~\ref{sec:methods} using three corpora of text that differ in linguistic complexity. The goal is to validate the effectiveness of these metrics and justify their use in our main analyses. Specifically, we assess: (1) whether the lexical divergence measure used in Section~\ref{sec:results_2} can reliably distinguish between texts of different complexity, and (2) whether the lexical scores defined in Section~\ref{sec:results_3} can meaningfully capture such differences—i.e., whether these scores vary significantly across text types. This validation ensures that our lexical metrics are both robust and appropriate for the analyses carried out in the main text.

    For this analysis, we construct three corpora of equal length ($6.2 \cdot 10^4$ words each), representative of different levels of linguistic complexity:
    
    \begin{itemize}
        \item[1.] \textbf{Learners}: a corpus built from the COREFL (Corpus of English as a Foreign Language) dataset~\cite{lozano2020designing}, composed of texts written by non-native English speakers. We consider this corpus a benchmark for low linguistic complexity.
        
        \item[2.] \textbf{Simple}: a corpus assembled by combining three classic children’s books with intentionally simple and accessible language: \textit{Grimm’s Fairy Tales} by the Brothers Grimm, \textit{The Wizard of Oz} by L. Frank Baum, and \textit{Aesop’s Fables}. All texts were retrieved from Project Gutenberg~\cite{gutenberg}. This corpus serves as a benchmark for moderate complexity.
        
        \item[3.] \textbf{Complex}: a corpus composed of three literary works known for their lexical and syntactic complexity: \textit{Ulysses} by James Joyce, \textit{The Sonnets} by William Shakespeare, and \textit{The Iliad} by Homer (in English translation). These texts were also obtained from Project Gutenberg~\cite{gutenberg}. This corpus serves as a benchmark for high complexity.
    \end{itemize}
    
    To faithfully replicate the analytical procedure employed in the main text, we construct synthetic users analogous to Twitter users by assigning them short excerpts from the corpora, treated as artificial tweets. Each synthetic tweet is generated by randomly sampling a sentence of 15–25 words from one of the three corpora, reflecting the typical tweet length observed in our dataset (average: $\sim$17.5 words). Each user is assigned 1500 synthetic tweets sampled exclusively from a single corpus, ensuring that every user consistently exhibits one type of language style. In total, we generate $N_{\text{learners}} = 20$ users associated with the learners corpus, $N_{\text{simple}} = 20$ users with the simple corpus, and $N_{\text{complex}} = 20$ users with the complex corpus. These numbers are consistent with the typical community size observed in the empirical dataset (average: $\sim$21 users per community). In this way, we construct three artificial communities—learners, simple, and complex—each composed of users producing language with distinct levels of complexity.

    We then apply the procedure described in Section~\ref{sec:methods}. After removing stopwords and lemmatising the synthetic tweets of each user, we identify and exclude the kernel lexicon, defined as the 1000 most frequent tokens across all corpora. We adopt the same subsampling strategy used in the main text to reproduce the results in Section~\ref{sec:results_2} and Section~\ref{sec:results_3}: we sample 60\% of users from each community and compute the lexical distance $\Delta_{cc'}$ between pairs of communities, for a total of 1000 iterations. In each iteration, we also generate two random communities of size $N=20$ by sampling users across all groups. This simulates the null model used in the main text, where the underlying community structure (learners, simple, complex) is removed by random mixing. The lexical distance between these randomised communities provides a baseline for comparison.

    Figure~\ref{fig:test_lexical_distances} shows the distributions of lexical distances across 1000 subsampling iterations. All empirical distributions (learners–simple, simple–complex, learners–complex) are clearly separated from the null distribution, with learners–complex showing the greatest divergence, as expected. $p$-values reported in the legend confirm the statistical significance ($p < 0.001$ in all cases).
    Notably, the lexical metrics span relatively narrow ranges. This is due in part to the exclusion of the kernel lexicon and in part to the subsampling procedure, which reduces variability. In this particular experiment, variability is further reduced by the relatively small size of the corpora and the homogeneity of the users within each group.

    We finally evaluate the lexical scores introduced in Section~\ref{sec:results_3}. For each user, we compute their vocabulary size ($S_i$), complexity ($H_i$), and completeness ($J_i$). We then apply the subsampling procedure: at each of the 1000 iterations, we sample 60\% of users from each community, compute the average score for each metric, and repeat the same for two randomly constructed communities of size $N = 20$. The resulting mean values, with confidence intervals, for learners, simple, and complex communities are reported in Table~\ref{tab:test_lexical_scores}.
    
    To assess whether these metrics effectively distinguish between communities, we compute the differences in average scores between each pair of communities ($\Delta S$, $\Delta H$, and $\Delta J$), and compare them to the null distributions of differences ($\Delta S_0$, $\Delta H_0$, $\Delta J_0$) derived from the random communities. This allows us to test whether observed differences exceed those expected by chance. The results, including $p$-values, are shown in Table~\ref{tab:test_lexical_scores}, confirming that the proposed metrics are sensitive to linguistic differences across texts of varying complexity.

\section{Results with Louvain clustering}
\label{app:louvain}

    In this section, we replicate the analyses presented in the main text using an alternative community detection method—namely, the Louvain algorithm~\cite{blondel2008fast}, in place of the hierarchical Stochastic Block Model (hSBM) described in Section~\ref{sec:methods}. To ensure robustness, we run the Louvain algorithm 1000 times and retain the partition that yields the highest modularity score.
    
    The results closely mirror those obtained with hSBM, indicating that the relationships observed between network structure and lexical metrics are not an artefact of the specific community detection method employed.

    \begin{figure*}[tbh]
        \centering
        \includegraphics[width=\linewidth]{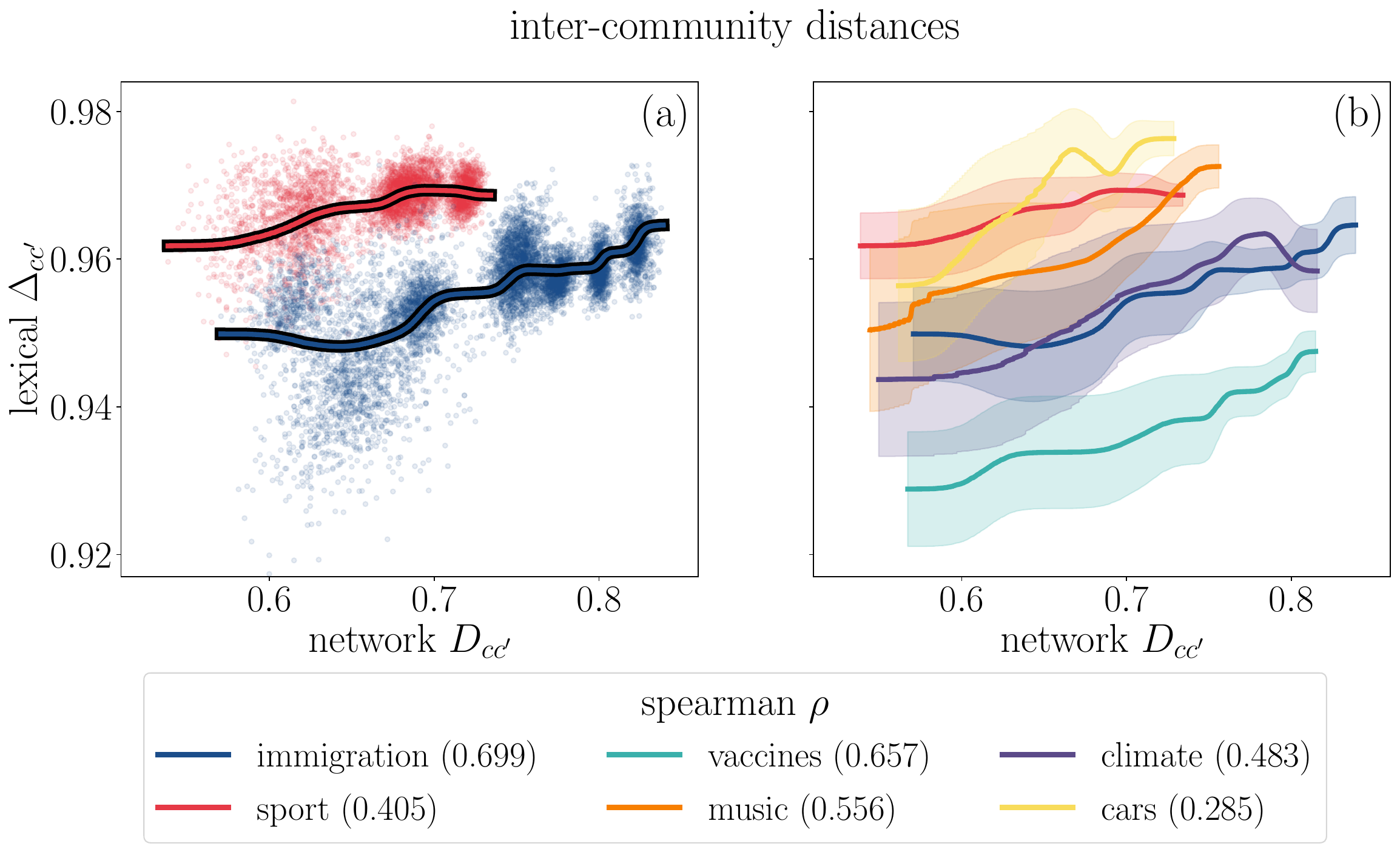}

        \caption{\textbf{Lexical divergences across communities correlate with distances on retweet networks (Louvain algorithm).} 
        \textbf{(a)} Inter-community lexical distance plotted as a function of network distance. Each point represents a pair of communities, subsampled using the procedure described in Section~\ref{sec:methods}. Solid lines indicate the median trend. 
        \textbf{(b)} Summary of results across all six topics. Solid lines denote the median values obtained through subsampling, while shaded areas represent the interquartile range. All topics exhibit a significant correlation, as measured by the Spearman correlation coefficient $\rho$, with statistical validation provided by null model tests (Table~\ref{tab:null_louvain}).}
        
        \label{fig:fig2_louvain}
    \end{figure*}

    \begin{figure*}[tbhp]
        \centering
        \includegraphics[width=0.8\linewidth,]{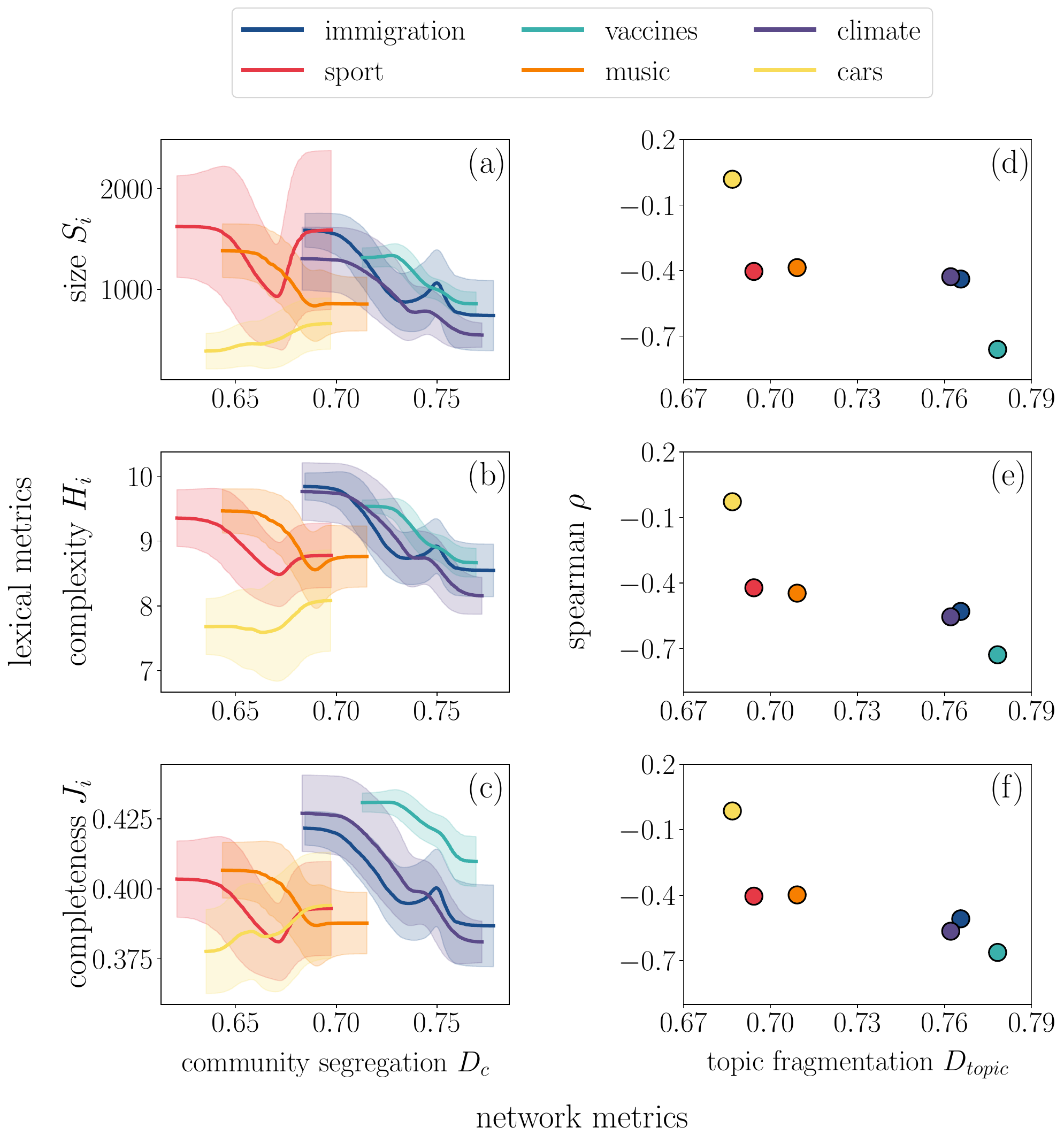}

        \caption{\textbf{Lexical scores as a function of community segregation and topic fragmentation in the retweet network (Louvain algorithm).}
        \textbf{(a–c)} Lexical metrics plotted against the community network segregation $D_c$ for all six topics. Solid lines indicate the median trend across subsamples, while shaded areas represent the interquartile range. Topics with higher overall segregation (cool colours) display stronger correlations, whereas less segregated topics (warm colours) show weaker—but still statistically significant—associations. 
        \textbf{(d–f)} Spearman's $\rho$ correlation coefficients between lexical scores and community segregation, shown as a function of topic fragmentation $D_{topic}$ (see Methods). The plots confirm that the strength of the correlation increases with topic fragmentation.}

        \label{fig:fig3_louvain}
    \end{figure*}

    \begin{table*}[t]
    \begin{tabular}{|c|c|c|c|c|c|c|c|c|c|c|c|c|}
    \hline
     & \multicolumn{3}{c|}{\textbf{lexical distance}} & \multicolumn{3}{c|}{\textbf{vocabulary size}} & \multicolumn{3}{c|}{\textbf{complexity}} & \multicolumn{3}{c|}{\textbf{completeness}} \\
    \hline
    \textbf{topic} & $\rho_{\text{data}}$ & $\rho_0$ & $p$ & $\rho_{\text{data}}$ & $\rho_0$ & $p$ & $\rho_{\text{data}}$ & $\rho_0$ & $p$ & $\rho_{\text{data}}$ & $\rho_0$ & $p$ \\
    \hline
    immigration & 0.699 & -0.246 & 0.014 & -0.439 & 0.475 & 0.050 & -0.531 & 0.427 & 0.046 & -0.508 & 0.422 & 0.058 \\
    \hline
    vaccines & 0.657 & -0.021 & 0.002 & -0.761 & 0.487 & 0.007 & -0.729 & 0.450 & 0.008 & -0.662 & 0.338 & 0.017 \\
    \hline
    climate & 0.483 & -0.418 & 0.058 & -0.428 & 0.478 & 0.062 & -0.555 & 0.491 & 0.042 & -0.565 & 0.482 & 0.052 \\
    \hline
    sport & 0.405 & -0.357 & 0.025 & -0.404 & 0.503 & 0.003 & -0.422 & 0.469 & 0.008 & -0.404 & 0.478 & 0.013 \\
    \hline
    music & 0.556 & -0.195 & 0.001 & -0.386 & 0.362 & 0.001 & -0.447 & 0.295 & 0.003 & -0.399 & 0.329 & 0.004 \\
    \hline
    cars & 0.285 & -0.369 & 0.017 & 0.018 & 0.481 & 0.035 & -0.028 & 0.431 & 0.069 & -0.014 & 0.471 & 0.044 \\
    \hline
    \end{tabular}

    \caption{\textbf{Null model tests (Louvain algorithm).}
    Statistical validation of the results shown in Figure~\ref{fig:fig2_louvain} and Figure~\ref{fig:fig3_louvain}. The null model procedure follows the approach described in Section~\ref{sec:methods}. For each case, the empirical Spearman correlation coefficient $\rho_{\text{data}}$ is compared to the distribution of coefficients $\rho_0$ obtained under the null model. The resulting $p$-values confirm that the observed correlations are statistically significant ($p < 0.05$) in nearly all cases, with only a few exceptions that remain close to the significance threshold. These findings support the robustness of our results, even when adopting the Louvain algorithm for community detection.}

    \label{tab:null_louvain}
    \end{table*}

\end{document}